\documentclass[draft]{agujournal2019}
\usepackage{url} 
\usepackage{xr}
\usepackage[authoryear]{natbib}
\usepackage{amsmath}
\usepackage{wasysym}

\draftfalse

\journalname{GRL}
\graphicspath{{./}{Figures/}}
\begin{document}
\date{}


\title{Climate change alters teleconnections}

\authors{E. Vos\affil{1}, P. Huybers\affil{2}, E.Tziperman\affil{2}}

\affiliation{1}{Laboratoire de Météorologie Dynamique/IPSL, Sorbonne Université,
ENS,PSL Research University, Ecole Polytechnique, CNRS, Paris France.}

\affiliation{2}{Department of Earth and Planetary Sciences and School of Engineering and Applied Sciences, Harvard University, Cambridge, MA, USA.}
\correspondingauthor{Eran Vos}{eran.vos@gmail.com}

\begin{keypoints}
\item Anthropogenic climate change has altered large-scale climate teleconnections, affecting regional temperatures.

\item Changes to the teleconnections of ENSO, the IOD, and PDO between 1960--1990 and 1990--2020 are robustly observed.

\item Future climate change is projected to further alter teleconnection patterns, amplifying their regional influence.
\end{keypoints}
\begin{abstract}
Internal modes of climate variability, such as El Niño and the North Atlantic Oscillation, can have strong influences upon distant weather patterns, effects that are referred to as ``teleconnections''. The extent to which anthropogenic climate change has and will continue to affect these teleconnections, however, remains uncertain. Here, we employ a covariance fingerprinting approach to demonstrate that shifts in teleconnection patterns affecting monthly temperatures between the periods 1960–1990 and 1990–2020 are attributable to anthropogenic forcing. We further apply multilinear regression to assess the regional contributions and statistical significance of changes in five key climate modes: the El Niño–Southern Oscillation, North Atlantic Oscillation, Southern Annular Mode, Indian Ocean Dipole, and the Pacific Decadal Oscillation. In many regions, observed changes exceed what would be expected from natural variability alone, further implicating an anthropogenic influence. Finally, we provide projections of how these teleconnections will alter in response to further changes in climate. 
\end{abstract}

\section*{Plain Language Summary}

Large-scale patterns in the climate system, such as El Niño and the North Atlantic Oscillation, can have strong influences upon distant weather patterns, effects that are referred to as ``teleconnections''. However, it is unclear to what degree anthropogenic climate change is altering these teleconnections. We show that shifts in these climate connections between 1960--1990 and 1990--2020 cannot be explained by natural variability alone and can be attributed to human influence. We examine five major climate modes and find that their effects on remote regional temperatures have already changed, impacting many parts of the world. Finally, we provide projections showing how these climate teleconnections are likely to further evolve in a warming future.

\section{Introduction}

Modes of internal variability such as the El Niño-Southern Oscillation (ENSO), North Atlantic Oscillation (NAO), Southern Annular Mode (SAM), Indian Ocean Dipole (IOD), and the Pacific Decadal Oscillation (PDO) are recurring large-scale oceanic and atmospheric patterns. These patterns influence local and remote temperatures, precipitation, and extreme weather events \citep{karoly1989,wallace1981,alexander2002,hobeichi2024,casanueva2014,wang2003climate,yeh2018enso,reboita2021impacts,simon2015role,casselman2023teleconnection} with significant environmental, social, and economic impacts \citep{zebiak2015,wang2003climate,hardiman2020predictability,kurths2019unravelling}. Projecting changes to the effects of these modes in a warmer future climate is critical for adaptation efforts \citep{gillett2003climate,king2010roles,paeth2010enhanced}. 

Teleconnections are typically induced by atmospheric Rossby waves \citep{Hoskins-Karoly-1981:steady, hardiman2020predictability, reboita2021impacts, casselman2023teleconnection} that are triggered by anomalies of sea surface temperature or surface heat fluxes, often via the mediation effects of atmospheric convection. The teleconnection pattern of a given variability mode (for example, ENSO) can change between events because differences in the state of the ocean and atmosphere can affect wave propagation, including due to other modes that can exert a modulating effect on the teleconnections \citep{gershunov1998interdecadal, hardiman2020predictability, joshi2021role}. Previous studies \citep{yeh2018enso} found indications of changes in teleconnection patterns since the 1990s, including in the effect of ENSO on North American precipitation \citep{simon2015role} and the effects of the Indian Ocean warming on the NAO \citep{hoerling2004twentieth}. As climate change continues, such teleconnections may be further affected, as suggested for ENSO \citep{lieber2024historical, beverley2024drivers}. Some modes and the relation between them may change in the future, as projected for the NAO and AO under extensive climate warming \citep{Hamouda-Pasquero-Tziperman-2021:breakdown}.

Observational analyses, detection–attribution studies, and climate model projections indicate that anthropogenic forcing is modifying not only the mean state of the climate system but also the structure, strength, and spatial patterns of climate teleconnections. Early attribution work emphasized that spatial patterns in the response to anthropogenic forcing are subject to substantial uncertainty \citep{jones2016uncertainties}, yet more recent studies show growing evidence that teleconnections are being reshaped by anthropogenic climate change  \citep{bilbao2019attribution, deser2017forced}. For example, changes in the Walker circulation and tropical convection under warming have already altered ENSO variability and its global teleconnections, with both historical amplification and robust end-of-century increases in ENSO-driven precipitation anomalies emerging across models \citep{Power2013, cai2023anthropogenic, xie2010global}. Similarly, global-scale analyses demonstrate that warming reorganizes atmospheric bridges and storm-track pathways through which variability modes influence remote regions \citep{burke2017global}. Outside the tropics, Arctic amplification and sea-ice loss have been linked to the evolution of Arctic–midlatitude teleconnection pathways, with causal-network approaches revealing modified dynamical links between high-latitude variability, the polar vortex, and midlatitude circulation patterns in recent decades \citep{galytska2023evaluating}. These studies suggest that anthropogenic forcing can influence both the amplitude and spatial expression of teleconnections, motivating a systematic evaluation of how forced changes compare to internal variability within the climate system. 

Attribution of systematic shifts in global teleconnection patterns via covariance matrices of monthly mean temperature has not, to our knowledge, been addressed in the literature. Below, we first apply a covariance fingerprinting framework, projecting observed changes in temperature covariance onto model-simulated patterns of covariance change under greenhouse gas forcing. This approach allows us to formally attribute changes in global teleconnection structure to anthropogenic climate change. We later apply a second, complementary attribution approach that is based on a point-wise multilinear regression analysis.

\section{Results}
\subsection{Global analysis of climate change impacts on Teleconnections of Internal Variability Modes}

\begin{figure}[tb!] \centering
 \includegraphics[width=\textwidth]{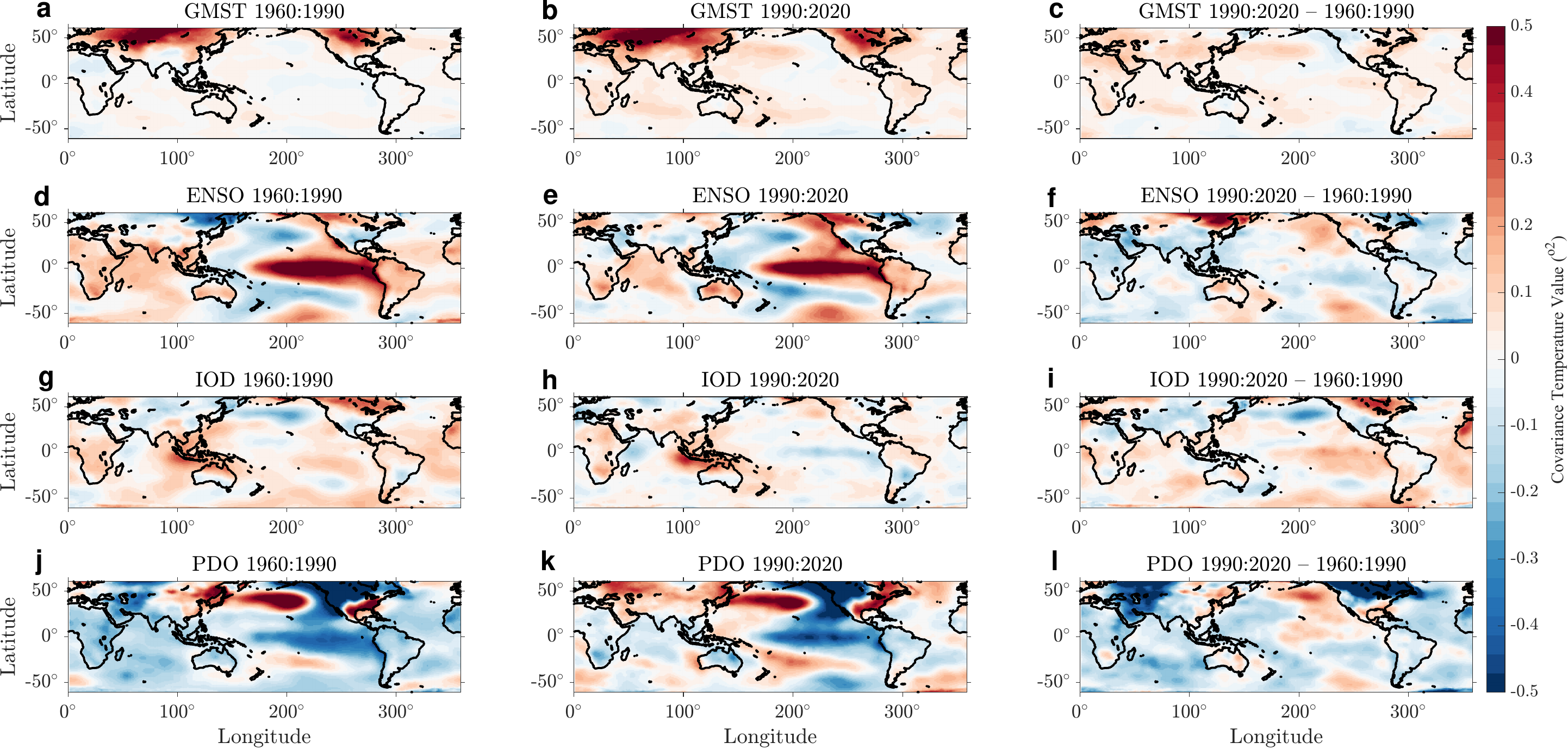}
 \caption{\textbf{Covariance changes across thirty-year epochs.} The left column shows covariance averaged over all grid cells (top row), the Niño3.4 region (second row), the IOD region (third row), and the PDO region (bottom row) for 1960--1990. The middle column is the same as the left column, but for the 1990--2020 period, and the right column shows the difference between the two periods. Note, the PDO covariance values in the bottom row are divided by 4 to facilitate the use of the same color range as the rest of the panels.}
     \label{f:Cov_recons}
 \end{figure}

We first examine the change in global teleconnection patterns using the covariance matrix of monthly-average temperature anomalies. Given $T_i$ as the monthly temperature anomaly at grid point $i$, the covariance matrix is $C=\{c_{ij}\}=\langle T_iT_j\rangle$ where angle brackets indicate time averaging. Therefore, row $i$ of the covariance matrix corresponds to the covariance of the monthly temperatures at the geographical location corresponding to this index with all other locations, and can be plotted as a map. We compose such covariance matrices of the monthly-average temperature anomalies among all grid cells between 60N--60S for the observations \citep{rohde2020berkeley} and the model control and historical runs \citep{CESM2_2020,Kay-Deser-Phillips-et-al-2015:community} (see Methods). To visualize the covariance change, we plot maps of the temperature covariance during 1960--1990 and 1990--2020, as well as the difference between them in Fig.~\ref{f:Cov_recons} as follows.  Note that the left and middle columns display most of the well-established teleconnections (discussed in the Supplementary Information), such as the positive correlation between ENSO and Australia. To focus on ENSO teleconnection changes, for example, we average the rows of the covariance matrix corresponding to all $i$ points within the NINO3.4 region and plot the resulting average as a map. Similarly, the changes to the IOD and PDO teleconnections are depicted as appropriate averages over the corresponding regions, as discussed in the Methods section (Eqn.~\ref{eq:plotting-covariance}). For the effect of GMST, we average all rows, weighted by cosine latitude. Because the covariance fingerprint is constructed from surface temperature covariances, SAM and NAO are excluded. Note that changes in teleconnections may include spatial shifts, changes in frequencies, amplitude, or the relation between variability modes. Our focus here is on the changes to spatial patterns (of monthly temperatures) and their amplitude as seen through the covariance matrix. 

Various changes in covariance are seen in the results plotted in the right column of Fig.~\ref{f:Cov_recons},  which shows the difference between the periods of 1990--2020 and 1960--1990. For example, Fig.~\ref{f:Cov_recons}f shows that ENSO teleconnections become stronger in Australia, as indicated by the red color there. Figs.~\ref{f:Cov_recons}d,e show that when El Niño events occur, the monthly temperatures over Australia increase. The panels on the right hand side (e.g., Fig.~\ref{f:Cov_recons}f) then show the change in the teleconnection. Teleconnection may change for two reasons. First, a change in the sensitivity to the remote influence, where a positive difference then implies that during the latter period, 1990--2020, the response of monthly temperatures to El Niño events is stronger than during the preceding period. Second, covariance differences that show an intensified pattern (e.g., for ENSO and Australia) could be because ENSO variability itself increased.  Covariance fingerprinting analyses do not separate between these two possible reasons for the change in teleconnection, but the regression analysis in the following section does provide this information, as discussed there.  

In some cases, the sign of the ENSO teleconnection changes, which clearly implies a change in sensitivity.  For example, temperatures switch from increasing with El Niño (positive teleconnection) to decreasing when El Niño occurs (negative) in Northeast Africa.  ENSO teleconnection signs also switch over South Asia and Western Europe (Fig.~\ref{f:Cov_recons}f). 

To briefly review some other teleconnections changes, covariance with GMST shows increased warming over Africa, Asia, South America, and Eastern North America relative to other parts of the world (Fig.~\ref{f:Cov_recons}c). For the IOD (Fig.~\ref{f:Cov_recons}~g,h,i), we see a change in the amplitude and structure of teleconnections in Australia, Africa, Asia, and the Middle East. Finally, for the PDO (Fig.~\ref{f:Cov_recons}~j,k,l), we observe a sign switch in Western Europe and Eastern North America, as well as a change in amplitude in Africa, Australia, and South America.

To test if changes in global teleconnection patterns are attributable to anthropogenic climate change, we use a covariance fingerprinting approach (see Methods) that is similar to the standard fingerprinting methods \citep{hasselmann1979signal, hegerl1996detecting,ribes2009adaptation} but involves projecting the observed change in covariance onto that obtained from climate model simulations forced by historical changes in greenhouse gasses, aerosols, land use, and solar variability. We calculate a scalar measure of this projection, referred to as $S_{\text{obs}}$, 
\begin{equation}
  S^{\rm obs} = \frac{\left\langle \Delta C^{\rm obs},\Delta C^{\rm model}_{\rm forced}\right\rangle_F}{\|\Delta C^{\rm model}_{\rm forced}\|_F^2}.
  \label{eq:S_obs}
\end{equation} 
$S$ quantifies how strongly the observed change in covariance projects onto the model’s forced pattern of changes in covariance, normalized by the square of the latter. The subscript $F$ refers to the Frobenius norm, or the square root of the sum of the squares of all elements. 
$S$ inherently contains all modes, including GMST. Intuitively, if the covariance matrices are identical, $S = 1$. If they share the same sign pattern, $S$ is larger than 0. Conversely, if the sign pattern is opposite, $S$ becomes negative. Thus, $S$ provides a scalar measure of the similarity between covariance matrices. 

\begin{figure}[tb!] \centering
 \includegraphics[width=0.5\textwidth]{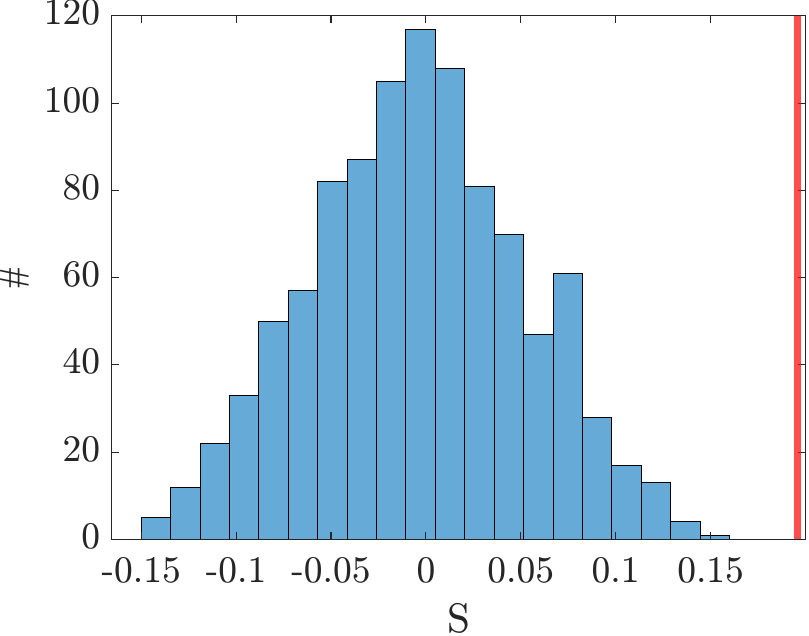}
 \caption{\textbf{Covariance fingerprint.}  $S_\text{obs}$ (red line) is larger than any of 1000 realizations of $S_\text{null}$ (histogram).  $S_\text{obs}$ is the observed covariance difference between 1960--1990 and 1990--2020 projected onto that produced by CESM2-LE historical runs. Each $S_\text{null}$ realization is the same as $S_\text{obs}$ but is computed using pairs of 30-year periods that are randomly drawn from the CESM2 long control run.}
     \label{f:fingerprint}
 \end{figure}

Fig.~\ref{f:fingerprint} compares $S_\text{obs}$ against 1000 realizations of $S_\text{null}$, a covariance fingerprint calculated between randomly selected pairs of 30-year periods in the CESM2 control run \citep{CESM2_2020,Kay-Deser-Phillips-et-al-2015:community}.  $S_\text{null}$ samples random changes in covariance due to natural variability. We find that all realized values of $S_{\text{null}}$ are smaller than $S_{\text{obs}}$. This calculation thus indicates that changes in global teleconnections are inconsistent with being solely the result of internal variability and, instead, are attributable to anthropogenic climate change.

The covariance fingerprinting method used here is a statistical analysis that tests for changes across all modes and locations and therefore captures the significance of the global change. It does not, however, provide regional specificity. The next section presents a complementary analysis that focuses on regional aspects.

\subsection{Spatially-resolved significance of changes in teleconnections}

Using the covariance fingerprinting approach, we find that global teleconnection patterns changed between 1960--1990 and 1990--2020 in response to anthropogenic climate change. In this section, we complement the above global analysis by examining point-wise changes to teleconnections for GMST and each of the five modes: ENSO, NAO, SAM, PDO, and IOD. Note that we now include the NAO and SAM, whose indices are defined in reference to sea-level pressure patterns, as opposed to temperature patterns. Our approach is based on multilinear regression (Methods, Eqn.~\ref {eq:multilinear-regression}), which examines the differences between 1960--1990 and 1990--2020 in relation to anthropogenic climate change by comparing observations to a large ensemble of model results \citep{Kay-Deser-Phillips-et-al-2015:community}. To stabilize the multilinear regression, the modes are orthogonalized and standardized (see Methods). 

\begin{figure}[tb!] \centering
   \includegraphics[width=\textwidth]{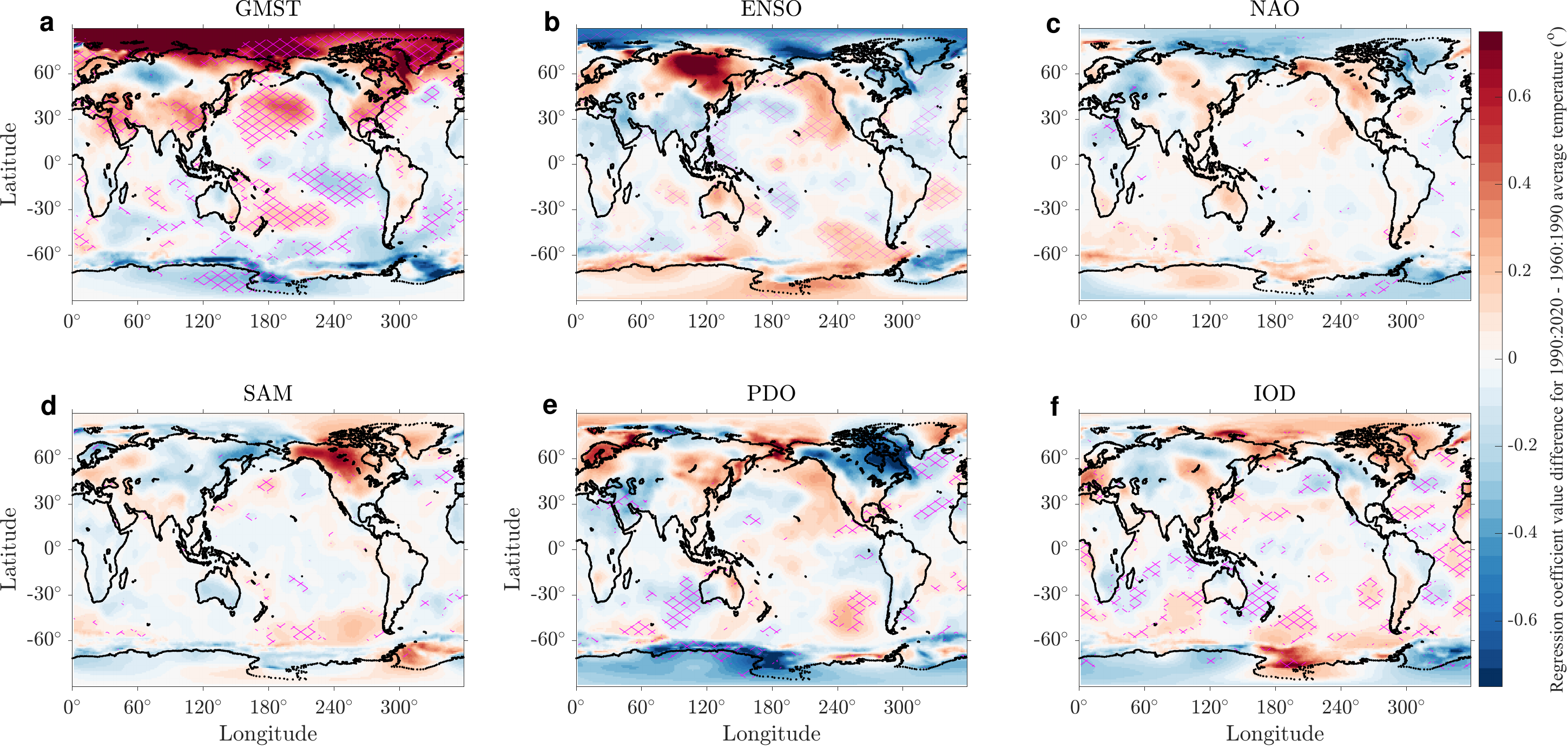}
   \caption{\textbf{Point-wise attribution.}  Differences in regression coefficients between 1960--1990 and 1990--2020 for five internal variability modes and GMST.  Positive values indicate increased co-variation of monthly temperatures with the corresponding mode. Locations where the differences between the periods can be attributed to anthropogenic climate change are indicated by pink crosses.}
   \label{f:Tavg_diff_1960-1990_climate}
 \end{figure}

Before proceeding to a point-wise attribution analysis of the difference between the two periods, we examine the contribution of each mode and GMST to monthly average global temperature anomalies over 1950--2020 (Fig.~SI1), finding well-known patterns that are listed explicitly in the supplementary \citep{cai2011teleconnection, yang2012systematic, cai2020climate, anderson2017life, roy2016enso, zhou2024interannual, kumar2017north, sandler2024connection, linderholm2011interannual, shi2019simultaneous, saji2005indian}. Differences in regression coefficients between 1960--1990 and 1990--2020 are shown in Fig.~\ref{f:Tavg_diff_1960-1990_climate}. The patterns of the temperature-based modes (GMST, ENSO, and IOD shown in Fig. 3a, b, and f) share features in common with the covariance-based results (shown in Fig. 1c, f, and i), with pattern correlations of 0.77, 0.98, and 0.59, respectively.   In addition to the aforementioned changes in the response to GMST, we see strong Arctic amplification (Fig. 3a), indicating a greater response of monthly temperatures in this region to GMST during recent decades. The regression approach, unlike covariance fingerprinting, however, shows that GMST is associated with the well-known cooling pattern in the East Pacific (Kosaka \& Xie, 2013) and that ENSO is anti-correlation with Arctic temperatures.  Differences arise between our two approaches because the regression-based approach isolates the component of temperature variability attributable to each mode after removing shared influences among GMST, ENSO, and IOD. In contrast, the covariance maps reflect the full co-variability between regional and base-region temperatures, including common signals from inter-mode correlations.  Furthermore, the regression approach uses standardized amplitudes of each mode. 

We focus on regions where changes in regression coefficients are statistically significant, as tested via bootstrapping (Fig. SI4). These changes are not explained by internal variability in the CESM2 control run (P $<$ 0.05; Fig. SI4, SI7) and are consistent with those found in the CESM2-LE historical runs (P $>$ 0.05; Fig. SI6). In general, 5\% of grid points may be expected to pass a significance test at a 95\% significance level. The fraction of grid points that pass, however, is significantly higher for GMST (28\%), ENSO (20\%), PDO (9.8\%), and IOD (15\%) teleconnection changes (Fig.~\ref{f:Tavg_diff_1960-1990_climate}). Values are more comparable to those expected by chance for the NAO (3.4\%) and SAM (4.0\%), reflecting primarily that the observed changes do not correspond with those produced by the CESM2 historical run. Note that for the NAO and SAM, large portions of the domain do not show statistically significant differences between the two periods when evaluated against the bootstrapped data.

Returning to the two reasons that may lead to teleconnection changes, Supplementary Table SI2 shows the variance of each index used in the regression analysis for each of the two periods. While the GMST variance has increased as expected due to the acceleration of warming in the latter period, the variance of other modes is basically unchanged. Note that the PDO time scale is decadal, and a 30-year period is insufficient to obtain a robust estimate of its variance. Based on these results, one may conclude that the teleconnection changes presented here and in the previous section arise not from changes in the modes themselves (e.g., ENSO), but from changes in the atmospheric medium through which these teleconnections propagate. This interpretation is supported by the fact that the time series of each mode (Supplementary Fig. SI10) shows no major shifts in behavior over the past three decades.

\subsection{Projected changes in teleconnections}

\begin{figure}[tb!] \centering
   \includegraphics[width=\textwidth]{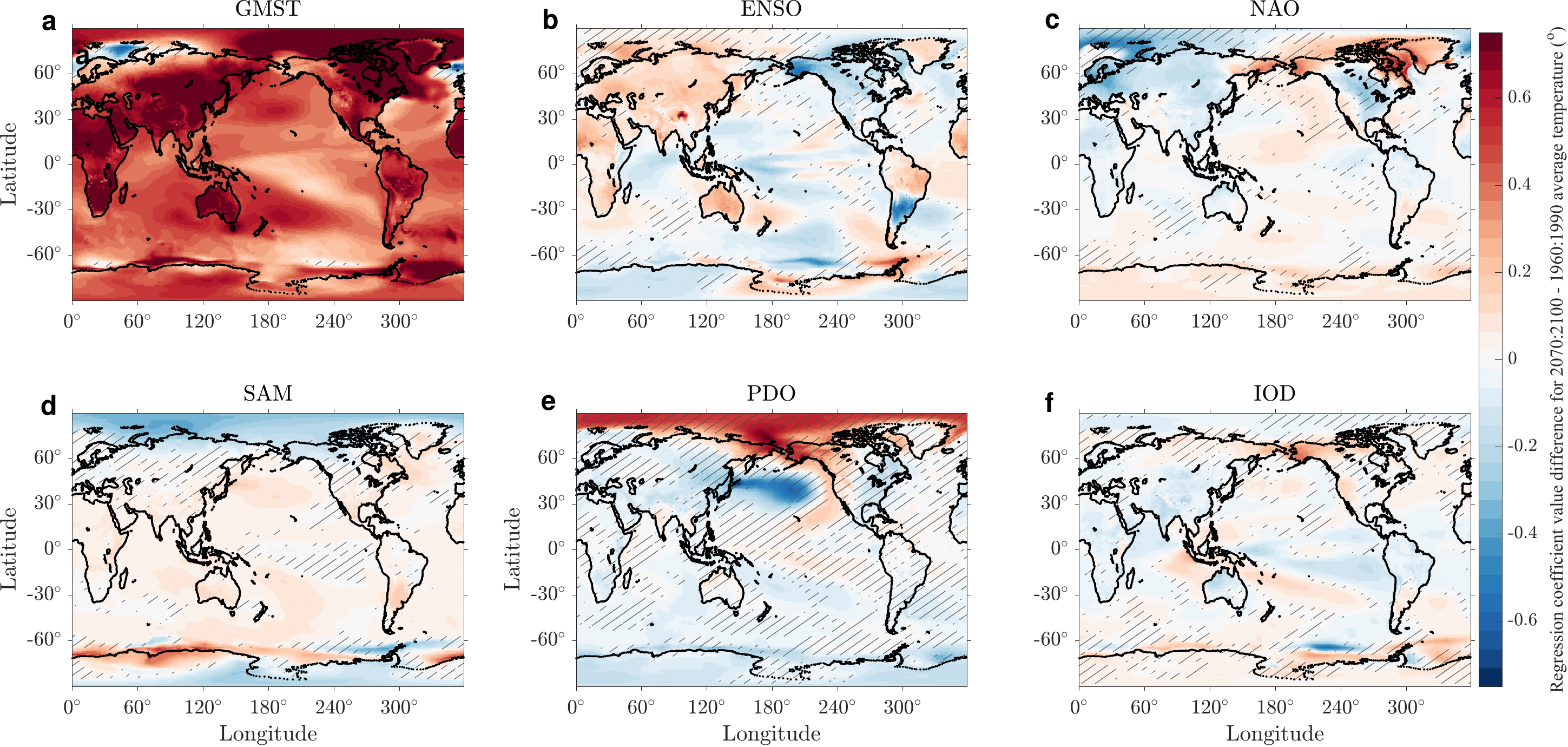}
   \caption{\textbf{Climate change projections.} Similar to Fig.~3, but for the difference in regression coefficients between 1960-1990 and 2070-2100 using simulations.  Temperatures and mode indices are from CESM2-LE simulations. In this case, locations where results are statistically insignificant are indicated by gray hatching.}
   \label{f:Tavg_diff_2070-1960}
\end{figure}

Finally, we examine how the teleconnections of these modes may further change in a warmer future climate by applying our multilinear regression analysis to the CESM2 large ensemble \citep{Kay-Deser-Phillips-et-al-2015:community} under the RCP 8.5 scenario \citep{riahi2011}. Maps of regression coefficient differences between 1960--1990 and 2070--2100 for the mean of the large ensemble are shown in Fig.~\ref{f:Tavg_diff_2070-1960}. Hatched regions represent locations where results are statistically insignificant (the 5--95 percentile range calculated over all ensemble members includes the zero value; note that this hatching represents a different meaning from that in Fig.~\ref{f:Tavg_diff_1960-1990_climate}, hence the different hatching pattern and color used; see a complementary presentation of the same results in Supplementary Fig.~SI15). The effects of GMST on monthly temperature increases, reflecting that a greater rate of warming is anticipated under RCP 8.5 in 2070--2100 relative to 1960--1990. We use the extreme RCP 8.5 scenario to maximize signal-to-noise ratio, but further examination of teleconnection changes for more modest---and perhaps more realistic---warming scenarios is warranted.

Projections of changes to ENSO teleconnections (Fig.~\ref{f:Tavg_diff_2070-1960}b) show amplification of some observed trends over the past decades, but the pattern of change is neither uniform nor simple. Positive teleconnections in Australia and Northern South America become stronger, as does a negative teleconnection in Southern South America (Figs.~SI11--13). The pattern over much of the Americas, however, generally weakens or changes. Other mode changes can be seen in the other panels of Fig.~\ref{f:Tavg_diff_2070-1960}. We also note that the effects of the NAO on monthly temperatures become less strong over Europe, hence, the difference is negative; that the PDO dipole structure in the North Pacific strengthens, and that the influence of the IOD dipole on monthly temperatures weakens locally as well as in South Asia and North Africa.

\section{Summary and Discussion}

We find statistically significant changes to teleconnection patterns of the major internal climate variability modes using two complementary attribution approaches. The monthly temperature-based covariance was found to change between 1960--1990 and 1990-2020 based on a covariance fingerprinting approach. To verify the robustness of these results, we applied two formulations: one that captures both the spatial pattern and amplitude of the covariance matrix (Fig.~2), and an alternative formulation that evaluates the matrices in a manner more analogous to a correlation (Fig.~SI17). Both approaches yield consistent outcomes, capturing the historical changes in the covariance structure. Results are confirmed and extended to modes that are defined based on sea level pressure rather than temperature (NAO and SAM) using a point-wise attribution approach. This point-wise analysis suggests that ENSO teleconnections were altered by anthropogenic climate change in more than 20\% of global grid cells. Similarly, IOD teleconnections have changed in over 15\% of grid locations.  Some teleconnections that pass the statistical significance test, such as the IOD effects on the Northern Hemisphere---would benefit from a more mechanistic understanding of the change to increase confidence in their patterns, although equivalent effects of the IOD were pointed out previously \citep{annamalai2007possible}. Forced climate simulations that are found to capture historical changes (Fig.~2,3) indicate that monthly temperature covariations will continue to change under a warming climate (Fig.~4).   This highlights that the effects of climate change on temperature go beyond global or regional averages and also alter the patterns of internal variation across most of the globe.

\clearpage\newpage

\bibliography{Earth}

\newpage

\section*{acknowledgments}
   T was funded by the Department of Energy (DOE) Office of Science Biological and Environmental Research grant DE-SC0023134 and the Harvard Dean’s Competitive Fund for Promising Scholarship, and thanks the Weizmann Institute of Science for its hospitality during parts of this work.
   EV was supported by the Council for Higher Education of Israel, under the project "Integrating Climate Dynamics, Clouds, and Extreme Events through Teleconnections in Climate Networks.

\section*{Author Contributions}
E.V., P.J.H., and E.T. designed the research, participated in the analysis and the writing, and E.V. performed the calculations.

\section*{Competing Interests}
The authors declare no competing interests.

\section*{Data availability}

The Niño 3.4 time series is available here: https://psl.noaa.gov/data/timeseries/monthly/Niño34. \\
The NAO and SAM time series are available here: 
https://psl.noaa.gov/data/20thC\_Rean/timeseries. \\
The PDO time series is available here:
https://psl.noaa.gov/pdo/ \\
The IOD time series is available here:
https://psl.noaa.gov/data/timeseries/month/DMI/ \\
The SAM time series is available here:
https://psl.noaa.gov/data/20thC\_Rean/timeseries/monthly/SAM/ \\
The GMST time series is available here:
https://www.ncei.noaa.gov/products/land-based-station/noaa-global-temp \\
Post-processed data of CESM2-LE control run, historical run, and future RCP 8.5 run are available at
https://www.earthsystemgrid.org/dataset/ucar.cgd.cesm2le.output.html

\end{document}